# The microbiome science of composting and human excrement composting: a review

Jeff Meilander[1,2] and J. Gregory Caporaso[1,2,3]

1. Department of Biological Sciences, Northern Arizona University, Flagstaff, AZ, USA
2. Pathogen and Microbiome Institute, Northern Arizona University, Flagstaff, AZ, USA
3. Corresponding author: greg.caporaso@nau.edu

## Abstract

Linear waste management systems are unsustainable and contribute to environmental degradation, economic inequity, and health disparities. Among the array of environmental challenges stemming from anthropogenic impacts, the management of human excrement (human feces and urine) stands as a significant concern. Over two billion people do not have access to adequate sanitation resulting in a global public health crisis.

Composting is the microbial biotechnology aimed at cycling organic waste, including human excrement, for improved public health, agricultural productivity and safety, and environmental sustainability. Applications of modern microbiome -omics and related technologies have vast capacity to support continued advances in composting science and praxis. In this article, we review literature focused on applications of microbiome technologies to study composting systems and reactions. The studies we survey generally fall into the categories of animal manure composting, food and landscaping waste composting, biosolids composting, and human excrement composting. We review experiments utilizing microbiome technologies to investigate strategies for enhancing pathogen suppression and accelerating the biodegradation of organic matter. Additionally, we explore studies focused on the bioengineering potential of microbes as inoculants to facilitate degradation of toxins such as pharmaceuticals or per- and polyfluoroalkyl substances (PFAS). The findings from these studies underscore the importance of advancing our understanding of composting processes through the integration of emerging microbiome -omics technologies.

We conclude that work to-date has demonstrated exciting basic and applied science potential from studying compost microbiomes, with promising implications for enhancing global environmental sustainability and public health.

# Introduction

Exponential human population growth has profoundly disrupted Earth's natural cycles, resulting in climate change, biodiversity loss, and widespread pollution. Among the array of challenges stemming from anthropogenic impacts, the management of human excrement (HE; defined here as human feces and urine), stands as a significant concern, exacerbated by the considerable volume produced globally. On average, each person defecates 125 grams of feces per day [1–3] totalling approximately 45.6 kg per person per year or 373 billion kg per year globally. Additionally, humans generate about 1.5 liters of urine per person per day which equates to 548 L per year or 4.5 trillion L per year globally [3, 4].

Access to adequate sanitation infrastructure to properly treat HE varies widely across the world [5–7]. In areas where properly functioning infrastructure exists, the process can use exorbitant amounts of energy. According to the U.S. Department of Energy, [8] wastewater treatment plants (WWTP) in the United States consume more than 30 terawatt-hours (TWh) of energy; a single plant expends 25–40% of its annual operating budget on energy alone. In 2020, 41.7 million metric tons of carbon equivalent were produced by the wastewater sector, marking a 13.3% increase since 1990 [9] and contributing to a new record high in global atmospheric carbon levels of 421.8 ppm [10]. Furthermore, every flush of a traditional toilet expels 2-6 gallons of potable water along with HE into sewers, where it mixes with stormwater runoff, industrial waste, and other contaminants before reaching WWTPs [11].

WWTP influent includes high levels of nitrogen, phosphorus, heavy metals, pharmaceuticals, "forever chemicals" such as per- and polyfluoroalkyl substances (PFAS), and other toxins. [12–17] Diverse chemical, biological, and physical processes are required to treat this complex mixture, yielding treated wastewater suitable for recycling (to the environment, aquifers, or for industrial, recreational, or agricultural uses) as well as sewage sludge that can be repurposed through land application (agriculture or environmental remediation) [18–20], incinerated, landfilled, or further processed into biosolids. However, challenges remain in completely eliminating persistent environmental pollutants, especially PFAS, from biosolids which contaminate soils, ground water, and livestock [17, 21–23].

Biosolids are nutrient-dense organic materials derived from the stabilization of sewage sludge through processes such as aerobic or anaerobic digestion, composting, or thermal drying [24]. These treatment methods can transform the sludge into rich compost and fertilizers, facilitating its application in sustainable agricultural practices. The conversion of waste into biosolids contributes to a circular economy by cycling organic matter back into the environment, thereby enhancing soil health and promoting sustainable waste management [25, 26]. However this process requires significant infrastructure, water, and energy inputs [27, 28].

Composting toilets (CTs) offer a decentralized approach to HE management that contrasts sharply with traditional biosolids composting. By avoiding the introduction of HE to the wastewater stream at the source, CTs reduce the burden on WWTPs, thereby decreasing the need for water and energy-intensive treatment processes and avoid contamination with industrial wastes. HE treatment is achieved through composting, a human-managed and microbially-driven process, which facilitates natural decomposition cycles resulting in a safe, nutrient-rich, humus-like product [27, 29–31]. The recovery and reuse of key nutrients in HE—specifically up to 91% of nitrogen (N), 83% of phosphorus (P), and 59% of potassium (K)—can effectively reduce reliance on synthetic fertilizers [16], thus enhancing soil fertility and agricultural yields [4, 32, 33].

Effective management of human excrement composting (HEC) is critical for ensuring both environmental safety and human health. Effective composting begins with the proper loading of CTs with a balanced mixture of HE and bulking materials (BM), such as sawdust, straw, or wood chips. BM provides structure, absorbs excess moisture, and creates air pockets that enhance aerobic conditions, which are crucial for efficient microbial decomposition [27, 29, 31, 34]. Once the CT reaches capacity, the material should be transferred to an actively managed compost pile. During the thermophilic phase, where temperatures can exceed 55°C, a succession of microbial communities degrades organic matter and significantly reduces pathogen levels. Thermophilic composting has also been shown to degrade various pharmaceuticals present in HE [15, 35].

Concerns surrounding the utilization of biosolids and manure as soil amendments have historically revolved around mitigating pathogens [36–39]. However, recent attention has broadened to encompass additional issues including endocrine disrupting compounds (EDCs) [40], antibiotic resistance genes (ARGs) [41, 42], pharmaceuticals [15, 43–45] and heavy metals [12, 16, 46–48]. Unlike biosolids produced from WWTPs, CT-derived compost is expected to contain lower concentrations of pharmaceuticals, toxins, and PFAS since it is not mixed with industrial wastes.

Furthermore,the practicality and efficiency of CTs make them particularly suitable where traditional wastewater infrastructure is nonfunctional or nonexistent. Examples include research field sites, eco-villages, national parks (including backcountry areas without running water), and rural communities. CTs operate with minimal infrastructure, often without the need for plumbing, making them considerably cheaper and easier to install than WWTP-based systems [49–53]. This is especially beneficial in developing countries where the lack of existing sanitation infrastructure and the high costs associated with centralized systems make CTs an attractive and feasible option [27, 28, 54]. These systems also operate with little to no freshwater input, which is crucial in regions where water resources are limited. Additionally, CTs could play a pivotal role in off-Earth human settlements, where water conservation and nutrient cycling from HE almost certainly won't be optional [55, 56].

CTs also offer significant public health benefits. With an estimated 2 billion people worldwide lacking access to adequate sanitation facilities and 450 million resorting to open defecation, exposure to dangerous levels of potential pathogens is significant [7]. This exposure contributes significantly to diarrheal diseases, which stand as the second leading cause of mortality among children under five years of age (over 1,200 deaths per day globally) [57–59]. Furthermore, an estimated 1.8 billion individuals globally rely on pit-latrines as a fundamental sanitation solution [60]. However, these latrines entail inherent risks, including soil leaching, groundwater contamination, and removal and storage of the excrement within which raise significant concerns regarding the transmission of enteric pathogens [60–62]. By providing a sanitary alternative, properly managed CTs help reduce the spread of human-associated pathogens, thereby improving overall community health [63–67].

While CTs offer sustainable and circular approaches to nutrient management, to fully harness the potential of composting in addressing HE management challenges, it is crucial to improve our understanding of the HEC process. Since Antonie van Leeuwenhoek first observed microorganisms (which he referred to as "animalcules"), and the pioneering work of scientist including Snow, Pasteur, Semmelweis, Koch, and Fleming, the field has heavily relied on culture-based identification of microbes, a technique that has been extensively applied to composting research generally and HEC research specifically [27–29, 31, 64, 68]ver the past two decades, our ability to investigate microbiomes (i.e., complex communities of microorganisms), has benefited from major technological innovations [69–71], yielding diverse new insights into the microbial world and its impacts on human and environmental health [72–76]. These advancements have sparked a scientific revolution in microbiology, leading to a paradigm shift in our understanding of the diversity of life [75, 77–80], elucidation of new opportunities for bioremediation [81–84], the era of microbiome-based medicine [85–89], and advancements in composting and manure and waste management [54, 90–94].

The democratization of Next Generation Sequencing (NGS) and the increasing accessibility of microbiome multi-omics has paved the way for future opportunities in the bioengineering of microbial communities and their successional patterns. Early examples range from the prediction of microbiome assembly and community dynamics based on encoded carbon catabolism and GC count [95–98] [99]. Despite the recent and widespread application of these technologies in analyzing microbiomes across various habitats, including diverse composting scenarios and waste treatment systems [91, 92, 100, 101] there remains a notable gap in their application to HEC research.

In this review, we begin by providing a brief overview of the history of human waste management in Western societies. We then define composting, describing how the process works with references to composting scientific literature. We then illustrate insights that NGS and other -omics technologies have revealed regarding various composting and other waste treatment methods. We conclude by advocating for the need to apply these advanced technologies to bolster HEC research, highlighting areas where these technologies could be applied to advance our understanding and praxis of HEC to catalyze advancements in sustainable waste management practices.

## Brief History - Laying the Groundwork

For centuries, agrarian societies disposed of HE directly onto agricultural lands [102]. Later referred to as 'night soil,' it was often collected and transported under the cover of darkness due to the widespread aversion to visibly handling and transporting the excrement [102–104]. As civilizations expanded, waste removal systems became essential to removing HE from burgeoning city centers, and HE was typically directed towards agricultural fields or bodies of water [102, 105, 106]. Advanced drainage and piping systems were constructed by the Minoan and Harappan civilizations during the Bronze Age and the Romans later engineered the Cloaca Maxima, which diverted HE directly into the Tiber River [105, 106]. Cultural, social, economic, and political influences reinforced the importance of recycling night-soil but rapid population growth, population density increase, and industrialization after the scientific and industrial revolutions substantially increased volumes of HE that needed to be managed. [102–104, 107]. Eventually, transporting night soil to surrounding agricultural areas became cost-prohibitive, exacerbating the already formidable management challenges.

Flush toilets, first developed by Sir John Harrington in 1596 and later patented by Alexander Cummings in 1775 [103], introduced new challenges despite their convenience and growing popularity in Europe. The excess flush water often flooded cesspools, leading to the contamination of drinking and groundwater while also diluting the excrement, thereby reducing its value as fertilizer and complicating its removal and transport. [103, 108]

In London, these new challenges coupled with the lack of proper sanitation infrastructure culminated in severe public health crises, including multiple cholera outbreaks and the infamous "Great Stink" of 1858. These events underscored the dire need for better HE management and marked the beginning of what became known as the "Sanitary Awakening." During this period, significant advances in microbiology, led by scientists like Louis Pasteur and Robert Koch, expanded the germ theory of disease, further highlighting the critical importance of proper sanitation [109–111].

In response to the urgent need for improved sanitation, Sir Joseph Bazalgette, chief engineer of London's Metropolitan Board of Works, and Colonel William Haywood were tasked with constructing an extensive sewer system that diverted raw sewage from the city into the Thames River [110]. Although this project reduced the visible presence of sewage in public spaces, it still discharged waste into a primary source of drinking water. Years later, in 1913, Edward Ardern and W.T. Lockett developed the activated sludge process, a biological treatment method for sewage sludge which became widely adopted after World War I [112].

These engineering and scientific breakthroughs also led to a significant reduction in diarrheal diseases, improved life expectancy, and laid the foundation for modern sanitation practices. The impact of these advancements was so profound that Annabel Ferriman (2007) would later highlight them in a paper titled, "BMJ readers choose the 'sanitary revolution' as greatest medical advancement since 1840," emphasizing how sanitation, germ theory, and bacteriology fundamentally changed societal approaches to waste treatment [113].

Despite technological, scientific, and engineering advancements over the last century, we continue to face monumental human and environmental health challenges related to HE. HEC and CTs present a multifaceted solution that not only reestablishes closed-loop nutrient cycling but also increases energy and water conservation, reduces environmental pollution, and improves the quality of life for many communities [48, 114, 115]. However, transitioning to these innovative treatment methods involves navigating complex cultural, economic, and technological barriers. For example, cultural preferences for traditional flush toilets and the perception of CTs as unconventional or less convenient hinder their acceptance [116–119].

Regardless of the challenges, advancing scientific understanding of HEC is essential for facilitating the broader adoption of HEC and CTs. Developing a deeper understanding of composting fundamentals and enhancing this knowledge through advanced microbiological research will be crucial for overcoming adoption barriers. By utilizing the latest sequencing technologies and research, we can address the complexities of HEC, demonstrate its immediate benefits, and drive the necessary paradigm shift toward sustainable waste management solutions.

## Composting science

The escalating scarcity of landfill space [120] driven by the vast quantities of organic waste generated by modern throwaway societies [121–123], presents a significant environmental challenge. This issue is further compounded by the growing volume of livestock manure, particularly from cattle, pigs, and poultry, which has surged over the past two decades [124], contributing to environmental pollution and greenhouse gas emissions [125, 126]. Composting offers a viable solution by transforming these organic materials into nutrient-rich products that can be reintegrated into natural biogeochemical cycles, thus mitigating their environmental impact.

Composting has a rich history with references in the works of Shakespeare and statements of George Washington, the first president of the U.S.A. [127]. While there is no universally accepted definition of composting, it is generally described as a process that yields a stable, nutrient-rich end product by breaking down organic materials [128]. Unlike natural decomposition, the composting process is defined in part by the fact that it is actively managed. More comprehensive definitions emphasize the importance of aerobic conditions [29, 129, 130], the occurrence of specific microbial metabolic activities [131], and introduce the terms thermophilic (temperature range of 45-70°C) [31, 132, 133] and mesophilic (temperature range of 25-45°C) [133, 134] (which are used in this context to describe phases of the process where thermophilic or mesophilic microorganisms are most active). The U.S. Food and Drug Administration (FDA) [135] includes the need for specific time frames at designated temperatures, and Cesaro et. al (2015) [136] describes the importance of compost stability. Finally, Insam and de Bertoldi (2007) [137] clarify that composting relies on microbial communities for biodegradation ("the breakdown of a molecular structure into its elemental components"), rather than biotransformation ("the biological modification that alters the chemical structure").

For the purpose of this literature review, we have incorporated elements from various perspectives to define *composting* here as the "human-directed, microbially-driven process of aerobic biodegradation of organic wastes at temperatures between 45-70°C, yielding a soil amendment that is safe, mature (indicating a high degree of process completion), stable (resistant to further decomposition), and nutrient-rich." If the composting process does not reach temperatures between 45–75°C, it is classified as *mesophilic composting*. While mesophilic composting does not achieve the same rate or efficiency of pathogen degradation as thermophilic composting, prolonged treatment times have been shown to be effective for ensuring safety, though specific parameters leading to safe product are less well understood [66, 138].

Composting is a well understood process (Figure 1) that relies on compost operators to monitor and manipulate multiple variables such as carbon-to-nitrogen ratios (C:N), moisture, bulk density, pH, and oxygen content [27, 29, 31, 34, 139–141] throughout four distinct phases (Table 1). *Feedstocks* are the organic, foundational components of compost (such as food waste or manure) that are collected and mixed with BM to optimize the microbial environment by attaining the preferred ranges of composting variables. Properly managing these variables within targeted ranges stimulates the metabolic activities of microbial communities thus elevating temperatures into the thermophilic range and enhancing biodegradation and reducing pathogen levels [66, 142, 143]. To reduce pathogen abundance, the World Health Organization (WHO) recommends that HEC reach 50° C or higher for between seven and 30 days followed by a 2-4 month curing phase [144] and the US EPA similarly requires temperatures above 55°C for at least three days when producing class A biosolids [145]athogenic fungal communities were observed to decrease concurrently with lower C:N ratios over time while wood saprotrophs increased. This was likely due to a reduction in nutrient availability for the pathogens, but reinforces the need to properly manage compost variables [146].

Composting investigations have traditionally relied on techniques such as microbial culture which allow researchers to characterize their physiological and catabolic properties [147–150]. However, ATP content analysis [151], phospholipid fatty acid (PLFA) analysis [152], and denaturing gradient gel electrophoresis (DGGE) [148, 152–154] have also yielded crucial insights into the dynamics of compost microbiomes throughout different composting phases and in response to varying physicochemical conditions [137, 147].

For instance, Ryckeboer et al. (2003) [147] investigated compost physicochemical properties alongside microbial succession using growth media and gas chromatography. They observed fluctuations in mesophilic fungi abundance following thermophilic temperatures, while mesophilic and thermophilic cellulolytic bacteria exhibited a significant increase post-cooling linked to further organic decomposition. The survival of fungi, yeasts, and streptomycetes was noted, possibly as spores or reintroduced from microhabitats on the edges of the composting piles reinforcing the need for proper turning and mixing of the compost. The prevalence of yeasts during the mesophilic phase was attributed to the initial low pH conditions, as yeasts can thrive at lower pH levels compared to bacteria.

Failing to properly manage the preferred ranges of compost variables can lead to adverse effects, particularly for novice users [149]. For example, temperatures of 46°C combined with a pH below 6.0 can inhibit mesophilic bacteria [155] whereas low pH and high concentrations of fatty acids can reduce microbial activity of thermophilic bacteria [156, 157]. Finished compost with a C:N >30:1 added to corn fields immobilized soil nitrogen [158, 159] and low C:N ratios increased phytotoxicity as a result of ammonia volatilization [158].

Curing, a critical phase in the composting process, allows compost to mature and stabilize, thereby ensuring its safety and effectiveness for agricultural use. Stability is characterized by a reduction in microbial activity, while maturity reflects the compost's capacity to support plant growth, often evaluated through bioassays. Mature compost is more effective at suppressing plant disease [133]. Neher et al. (2017) [160] proposed alternative methods for assessing disease suppression, such as quantifying the concentrations of enzymes produced by biocontrol microbes that degrade chitin and cellulose, key components of fungal mycelial cell walls, including those of the pathogen *Rhizoctonia solani*.

Comprehensive literature reviews and in depth research have described the benefits of adding finished compost to soil. These include increased organic content [161] and water retention [47, 162]; improved physicochemical properties [29, 163, 164]; reduced bulk density [165]; reduced plant and soil pathogen loads [166]; and reduced the need for pesticide and fertilizer application [27, 33, 139]. All of these factors lead to increased agricultural productivity [68, 167–170].

Despite the broad scientific consensus of compost benefits in agriculture, Bass et al., (2016) [171] demonstrated that there was no effect of compost application on banana and papaya productivity, and other studies demonstrated negative effects of compost application [172, 173]. These results may be due to nitrogen immobilization from high C:N ratios [29], microbial resource competition [159, 174], nutrient loss to soil surface layers [171], or phytotoxicity [173]. Alternatively, these anomalous results may be due to insufficient management of composting variables, experimental design issues, local weather and climate effects, or even aspects of the microbiomes of soil, feedstock, or BM.

In recent years, amplicon sequencing methods including rRNA or ITS profiling, have increasingly been applied to study microbiomes of composting systems [147, 175, 176]. This shift presents an opportunity to expand our current understanding of the microbiology of composting. Integrating other -omics technologies – for example, to identify active genes and metabolic processes – can enable bioprospecting from compost systems with the potential to uncover novel enzymes, antibiotics, and other bioactive compounds with industrial and pharmaceutical applications [177]. The microbial diversity inherent to the composting process represents a largely untapped resource for bioengineering initiatives aimed at enhancing specific microbial functions, such as the degradation of recalcitrant compounds or the biosynthesis of biofertilizers [178]. Additionally, the identification and validation of biomarkers of compost maturity—including specific microbial taxa, enzymatic activities, and metabolite profiles—enable more precise and reliable assessments of compost quality, thereby optimizing its effectiveness as a soil amendment [179]. This convergence of bioprospecting, bioengineering, and biomarker discovery paves the way for advancing the sustainability and efficiency of composting, in the context of HEC and beyond.

In the next sections we review literature on the applications of microbiome-focused multi-omics technologies in human excrement and more general composting research.

# Studying composting with microbiome science technologies

Succession in microbial communities refers to the changes observed following alterations that create new microhabitats [180]. Microbes drive the composting process and their relative abundances, interactions, and successional patterns are influenced by compost variables, recipes, phases, and substrates [181], as well as more general factors influencing microbial communities including resource availability [182, 183], inter- and intraspecific competition [36, 37, 39, 183], and abiotic and biotic factors [94, 184, 185].

Specific taxonomic groupings of bacteria (e.g., *Bacillales* and *Clostridiales*) and fungi (e.g., *Eurotiales* and *Glomerellales*) dominate compost microbiomes at different times [94]. These patterns are associated with different stages of the composting process, leading to a dynamic interaction between microbial populations and composting conditions [186]. For example, experiments have documented phasic and temporal responses between fungal and bacterial communities during pig manure composting [94] (analogous to those seen during forest restoration [187, 188]) as well as notable differences in microbiome composition as a result of specific plant species as compost substrates [189] and total available organic material [190]. Furthermore, microhabitats within compost systems differ spatially, temporally, and in response to changes in pH, temperature, or byproducts secreted by other organisms in the community [36].

As the composting process progresses, management of composting parameters shape composting dynamics and are essential for optimizing the final product through optimization of the microbial activity. Because DNA sequencing and other -omics technologies can identify precise patterns of microbial succession and reduce biases arising from culture-dependent research [52, 191], and provide unprecedented insight into microbial activity, they drive the development of new knowledge of the organisms and activities involved throughout the composting process [148]. This information can be applied to identify specific bioindicators [66, 179, 192] of compost phases or phase changes, and organisms capable of degrading toxins, pathogens, and pharmaceuticals [12–15]. Ultimately, this can facilitate better (and potentially fully automated) compost management.

Amplicon sequencing, particularly targeting the 16S rRNA gene, has become a cornerstone technique for studying microbiomes due to its high throughput and cost-effectiveness. The use of microorganisms by humans, for example in fermentation [193, 194], energy production [195], and other bioengineering and biotechnological advancements [196–199], has a history pre-dating our awareness of the existence of microorganisms. Modern microbiome analysis technologies have only opened new doors in this domain by providing a pathway to the rational design or optimization of these applications [81, 179, 192]. We now turn to review ongoing research that is unraveling the under-explored microbial communities involved in composting diverse organic materials including animal manure, household waste, biosolids, and HE. Ultimately we believe that this knowledge can be applied to HEC and CTs to promote sustainable waste management.

## Manure Composting

Global livestock production has increased substantially since the middle of the 20th century, leading to a corresponding rise in the accumulation of livestock manure. In concentrated animal feeding operations (CAFOs), manure is often stored in lagoons, posing risks such as rupture [200], leaching of N and P into nearby water sources [125, 126, 200, 201], and the contamination of drinking water [202]. Such contamination can induce adverse health effects such as cyanosis in infants [203], spread ARGs into the environment [204], and contribute to eutrophication [125, 205, 206] leading to toxic algal blooms [207].

Utilization of nutrients in manure can be achieved through composting or direct land application [208]. However, pathogens such as *Salmonella, E. coli, Listeria, Staphylococcus, Klebsiella, Enterobacter, Serratia,* and *Campylobacter* in manure pose significant concern. Ensuring that temperatures remain above 55°C for at least three days [150, 209], or adding chemical agents such as peracetic acid [67], urea, or ammonia [210–212] can reduce the load of pathogenic bacteria, viruses, and bacteriophages, offering a potentially useful treatment of manure prior to land application.

Inoculating compost with specific bacteria, such as *Kocuria, Microbacterium, Acidovorax, and Comamonas,* could enhance compost quality and efficiency [150]. During co-composting of dairy manure and sugarcane leaves, an inoculant containing the thermophilic bacteria *Ureibacillus suwonensis, Geobacillus thermodenitrificans,* and *Bacillus licheniformis*, known to produce biosurfactants capable of increasing composting efficiency [213], was associated with a prolonged thermophilic phase and increased rate of organic matter degradation [214]. In another study [215], the addition of maize straw to chicken manure composting was followed by an increase in the abundance of thermophilic bacteria such as *Limnochordaceae, Planifilum,*

*Oceanobacillus*, and *Thermobifida* and an accelerated rate of early stage microbial succession compared to composting chicken manure alone.

In another study of composting microbiome succession that profiled both bacteria and fungi, in the initial phases of pig manure composting the bacterial taxa *Bacillales* and *Clostridiales* predominated, facilitated by abundant cellulose content. Subsequently, as the composting process advanced, fungal taxa such as *Eurotiales* and *Glomerellales* proliferated, correlated with the accumulation of recalcitrant compounds left behind as metabolic byproducts of bacterial decomposition, contributing to enhanced humification and further biodegradation [94]. The addition of fine coal gasification slag to the co-composting of wheat straw and pig manure led to the identification of fungal taxa that could reduce pathogens and increase the degradation of cellulose and lignin [216].

While not composting based on the definition we provided in this manuscript, Semenov et al. (2021) [217] observed an increase in microbial diversity and biomass in soils amended with cattle manure, likely due to the higher levels of total organic carbon and total nitrogen reported in the soil following the amendment [12, 218, 219]. Over the course of 44 weeks, amplicon-based detection revealed the immediate reduction of manure-associated bacteria, but spore forming organisms such as *Clostridioides*, *Paeniclostridium*, *Romboutsia*, and *Turicibacter* persisted nearly a year later. This suggests the importance of composting manure before direct application on food crop soils. *Acinetobacter* in particular, detected in high abundance at 8 weeks, raises concerns for safety due its antibiotic resistance properties [220].

By leveraging amplicon sequencing, research teams aim to further enhance manure composting efficiency [221] and engineer soil microbiomes [222]. We expect that advances in this area will accelerate sustainable food production [208, 223, 224].

## Food and Landscaping Waste Composting

In the US, 55% of municipal solid waste (MSW) in 2018 was disposed of in landfills. The composition of MSW included 24% food waste and 7% yard trimmings [123]. This represents an enormous volume of compostable materials that could be reallocated for economic and environmental benefits.

Using 16S analysis, Partanen et al. (2010) [225] investigated the microbial succession during aerobic composting of municipal organic waste blended with wood chips. They noted a significant prevalence of *Bacillus* spp. during the transition from the mesophilic to thermophilic phase, while *Actinobacteria* and *Thermoactinomyces* spp. were associated with rapid, oxygen rich composting conditions. In contrast, *Clostridioides* spp. were linked to low oxygen environments, despite the presence of thermophilic temperatures and elevated pH levels. Tortosa et al (2017) [131] explored the bacterial diversity in composting olive mill waste in Spain. They connected specific genera with composting phases and composting parameters such as pH, temperature, moisture, and humification.

Using compost derived from plant-based feedstocks, Pot et al. 2021 [101] found that different compost recipes had the greatest effects on the bacterial and fungal microbiome compositions relative to other variables including the degree of compost maturation (age) or optimization treatments (blending or sieving to increase organic matter content or acidifying to adjust pH). Matured compost (as compared to fresh, immature compost) and optimization treatments also had no significant effect on the carbon sources utilized by microbial communities across the different composts. However, a significant increase in plant growth promoting rhizobacteria (PGPR) such as *Nitrobacter* and *Pedobacter,* and plant growth promoting fungi (PGPF) such as *Trichoderma*, were observed as a result of maturation. Despite the lower temperatures (<40°C), human pathogen associated genera such as *Salmonella, Escherichia, Klebsiella, Shigella,* and *Enterobacter* as well as plant associated pathogens such as *Verticillium, Rhizoctonia, Fusarium, Pythium, Phytophthora, Sclerotinia*, and *Plasmodiaphora* were undetectable in plant-based, matured compost.

Amplicon sequencing can also suggest potential symbiotic or antagonistic relationships. For example, regardless of composting conditions, Bacillales increased concurrently with *Symbiobacterium* in the final stage of composting when organic matter was thoroughly degraded [226]. Another species of *Symbiobacterium* demonstrated an obligate commensal interaction with a thermophilic *Geobacillus* species in composts [227] and *Bacillus* and *Thermobifida* exhibited a synergistic interaction in the degradation of lignocellulose [228]. Dairy manure inoculated with competitive exclusion microorganisms including *Brevibacillus parabrevis, Pseudomonas thermotolerans,* and *Comamonas testosteroni,* isolated from various finished composts, demonstrated a reduction in virulent and avirulent strains of *E. coli* O157:H7 relative to non-inoculated samples [229]. McKellar et al. (2007) [39] also found a strain of *Pseudomonas fluorescens* in raw milk that demonstrated antagonism against *E. coli* O157:H7 due to the production of siderophores. *Pseudomonas* spp., therefore, could be useful biocontrol agents to reduce foodborne pathogens in compost. Similarly, *Comamonas testosteroni* was identified as a potential androgen degrader in aerobic sewage samples from a WWTP in Taipei, Taiwan [40].

As in manure composting research, studies of food and landscaping waste composting have suggested inoculants that may be relevant to the distinctive characteristics of the organic material. Niu and Li (2022) [228] used *Bacillus thermoruber* as an inoculant in food factory sewage sludge composting to improve lignocellulose breakdown, prolong the duration of the thermophilic phase, and increase composting efficiency. *Trichoderma*, along with bacterial species *Azospirillum brasilense* and *Agrobacterium* spp., were used as inoculants during the composting of straw, leaves, and grass, stabilizing the C:N ratio and increasing NPK [230]. These inoculants were undetectable in mature compost, but *Bacillus* (maintained throughout the composting process)*, Pseudomonas, Paracoccus, Chryseobacterium, Delftia,* and *Ochrobactrum* genera were identified. The diversity of the observed microbiome depended on environmental conditions, available feedstocks, and interactions among these factors. Additionally, Nakasaki et al. (2019) [226] utilized a yeast strain as an inoculant during the mesophilic phase to facilitate the degradation of organic acids in compost derived from food waste, rabbit food, and cooked rice. *Geobacillus* was also identified as an inoculant during the thermophilic phase of vegetable waste composting to increase microbial metabolism. [231].

## Biosolids Composting

WWT necessitates substantial financial investment, infrastructure development, material resources, and effective governance and regulatory frameworks to ensure the safe treatment and discharge of effluents. Biosolids, resulting from WWT, can undergo various disposal methods including landfilling, composting, incineration, and land application, while the treated wastewater can be reintroduced into receiving bodies of water or marketed as reclaimed water for applications such as irrigation, groundwater recharge, or industrial processes [24]. The chemical attributes of biosolids are significantly influenced by diverse waste sources and treatment methods, such as aerobic digestion, liming, or composting, thereby impacting their microbial compositions and pathogen abundance [232–235]. Consequently, researchers have utilized NGS techniques to expand our understanding of these microbial communities, aiming to enhance safety and efficiency of biosolid production.

Annually, global biosolids production exceeds over 33 million dry tons [236] 8 million dry tons of which are produced in the U.S. alone [237]. In 2022, 27% of biosolids generated in the U.S. were disposed of in landfills, thus reducing available landfill capacity [24]. Notably, Australia land-applies 83% of biosolids, while several

European nations apply over 60%, and the U.S. applies 50-55% [20, 238]. Land application of biosolids can improve the physicochemical properties of soils [239] and improve agriculture while displacing the use of synthetic fertilizers [240]. However, concerns regarding the land application of biosolids stem from the origin and composition of the influents through the contents of the effluents which can contain heavy metals, ARGs, pathogens, and EDCs.

Despite these concerns, biosolids represent a valuable nutrient source [239] typically containing between 50–90% of their N composition in organic compounds [232]. Biosolids also contain essential micronutrients such as B, Cl, Cu, Fe, Mn, Mo, and Zn which are crucial for plant growth but often lacking in conventional chemical fertilizers [241] and offer the potential to recover elements such as Ag, Cu, Au, P, Fe, Pd, Mn, Zn, Ir, Al, Cd, Ti, Ga and Cr [237].

Tracking microbial succession through the WWT process is essential for understanding the effectiveness of treatment methods, assessing potential risks to public health and the environment, and informing management practices to ensure the safety and sustainability of these systems. For example, seasonality and influent sources changed the biofilm formation in aerobic granular sludge treatment plants [242]. Using dewatered biosolids from a WWTP in Colombia, Bedoya et al. (2018) [243] noted a shift from *Pseudomonas* dominated biosolids during the dry season to *Coprothermobacter* dominated biosolids in the rainy season. *Pseudomonas* has been shown to tolerate elevated heavy metal concentrations and degrade xenobiotic pollutants and therefore could provide insight into bioengineering these traits [244].

An observed shift from methanogens to sulfur-metabolizing and surfactant-degrading bacteria in an anaerobic membrane bioreactor (AnMBR) [56] demonstrates the adaptability of microbial communities to challenges such as surfactant and sulfate presence in wastewater. Surfactant accumulation can impair treatment performance so by supporting bacteria capable of degrading surfactants and metabolizing sulfur, the AnMBR enhances its ability to manage complex wastewater compositions, improve treatment efficiency, and support sustainable water recycling. This technology's effectiveness in degrading organic pollutants and recovering nutrients, along with its low energy consumption, makes it particularly suitable for controlled ecological life support systems (CELSS) in space missions, thereby advancing space exploration capabilities.

Ye and Zhang (2013) [245] employed 454 pyrosequencing to assess bacterial diversity in influent, activated sludge, digestion sludge, and effluent from a WWTP in Hong Kong. In this system, seawater is used in toilet flushing and thus raises the salinity of the influent. They observed that activated sludge samples, utilized during aerobic WWT, represented a species-rich environment. The microbial composition of digestion sludge exhibited distinct characteristics relative to a freshwater sludge digester, including low evenness dominated by a high abundance of *Kosmotoga* species (approximately 67%). This abundance may be attributed to the incorporation of seawater but further studies are warranted to elucidate these relationships. Associating microbes or genes with parameters of different WWTP systems can ultimately inform more efficient management systems and protocols, or uncover organisms with biotechnological relevance, creating economic opportunities and environmental benefits [177].

Furthermore, identifying pathogen persistence through biosolids research is crucial for ensuring the safety and effectiveness of treatment processes and the subsequent reuse or disposal of biosolid materials. Bibby et al. (2010) [176] used 454 pyrosequencing to characterize pathogen diversity in biosolids subjected to mesophilic and thermophilic anaerobic digestion and composting. Their study consistently identified *Mycobacteria* and *Clostridia* across all samples, including opportunistic pathogens *M. forituitum*, *M. phlei*, and *M. chelonae*. In Ireland, a study utilized 16S and 23S sequencing to determine that Autothermal Thermophilic Aerobic

Digestion (ATAD), a tertiary treatment process for sewage, was capable of reducing pathogen and fecal coliform loads [246].

A different study conducted at a WWTP in Berlin, Germany, utilized full-length 16S sequencing to evaluate the WWTP's efficacy in mitigating pathogenic bacteria. Researchers noted increased levels of *Legionella* and *Leptospira* in effluent when compared to the influent [247]. If the effluent is not properly depleted of human pathogens, nutrient (C and N) loading from the effluent can increase pathogen abundance in the environment [248] and contribute to their residence in soil and water habitats [249]. Numberger et al. (2019) [247], however, were not able to detect enteric bacteria such as *Campylobacter* and *Salmonella* spp. in WWTP effluent but did observe a decrease in *Acinetobacter, Aeromonas, and Pseudomonas*.

Little et al. 2020 [250] identified *Devosia* and *Bradyrhizobium* genera, commonly associated with PGPB and phytohormone production, in freshly dewatered and 4 year old, stockpiled biosolids [251, 252]. Various *Bacillus* spp. were linked to the degradation of lignin and lignocellulose producing phenolic and quinone compounds. These compounds are crucial precursors for humus formation, and promote the maturation process in sewage sludge composting [228]. Additionally, fungal communities were found in mesophilic and thermophilic anaerobic digesters that likely catabolize organic matter, providing a valuable resource for archaeal and bacterial communities [253]. Other metabolic byproducts, such as volatile fatty acids and volatile phenolic acids may serve as potential nutrients for microbial groups [254].

During the thermophilic phase of ATAD, biodegradation of organic material resulted in a reduction in the activity of denitrifying and nitrifying bacteria likely attributed to an elevated pH [246]. *Desulfobacterales*, crucial in nitrogen cycling within mangrove forests through its contribution of dissimilatory nitrate reduction to ammonium (DNRA) genes, was prevalent in the influent of a WWTP [245] and holds potential for targeted utilization to enhance nitrogen cycling efficiency [255]. Other nutrient cycling bacteria such as *Comamonas denitrificans* (present in biofilms of activated sludge), *Nitrospira* spp*.* (nitrite oxidation), and *Simplicispira limi* (associated with enhanced biological phosphorus removal (EBPR)) to mitigate eutrophication [256] were identified in WWTP effluent [247].

Piterina et al. (2010) [246] observed increased levels of DNase activity, likely bolstered by abundant extracellular DNA stemming from cell lysis induced by thermophilic temperatures. This extracellular DNA could serve as a valuable nutrient source for other microbes within the community or as a vector for horizontal gene transfer (HGT) [257]. ATAD and thermophilic composting degrades exogenous DNA and therefore could mitigate antibiotic resistance through HGT, enhancing the safety of biosolids prior to their land application [100, 246].

Alternative treatments, such as pyrolysis and vermicomposting, offer promising avenues for biosolids management. Kimbell et al. (2018) [41] demonstrated that pyrolysis of biosolids reduces the abundance of ARGs released into the environment, reaching levels undetectable by quantitative polymerase chain reaction (qPCR) assays. Pyrolysis, as a thermochemical process, holds potential for converting biosolids into valuable products including pyrolytic oil, pyrolytic gas, and biochar [171, 258].

Additionally relevant to biosolids composting, utilizing annelids, particularly *Eisenia fetida*, in vermicomposting or vermistabilization, which involves stabilizing and reducing the volume and hazardous components of organic waste for environmental management purposes, has emerged as another valuable composting technique [259–265]. These techniques could be applied in composting toilets [266], biosolids management [267], and biowaste treatment [268] where the worms would play a pivotal role in the decomposition process, effectively reducing the mass of organic matter in the piles while simultaneously mitigating pathogen levels, degrading antibiotics, and reducing heavy metal content. Their feeding activity also accelerates the breakdown of organic materials, enhances microbial activity, and promotes aeration and moisture regulation within the composting matrix thus creating favorable conditions for beneficial microbial communities to thrive [259]. Moreover, the vermicomposting process (and resulting worm castings) has been shown to result in compost with improved physical, chemical, and biological properties compared to traditional composting methods, making it an attractive option for sustainable waste management [181]. Pre-composting biosolids prior to vermicomposting ensures thermophilic temperatures are reached to meet regulatory standards, and safeguards the worms as temperatures exceeding 35°C can be detrimental to their survival [269]. The by-products of pyrolysis and vermicompost not only offer potential applications in agriculture but also contribute to environmental restoration efforts.

## Human Excrement Composting

Launched in 2015 as part of the 2030 Agenda for Sustainable Development, the UN Sustainable Development Goals (SDGs) were created to address a wide range of global challenges, including poverty, inequality, climate change, environmental degradation, peace, and justice. They provide a blueprint for achieving a more sustainable and equitable future and were developed in response to growing recognition of interconnected global issues that require coordinated, long-term efforts by governments, businesses, civil society, and individuals. Goal 6 aims to ensure universal access to safe water, sanitation, and hygiene, improve water quality, protect water ecosystems, and promote sustainable water management by 2030 [270].

Achieving Goal 6 and advancing the reuse of human excrement as a soil amendment and fertilizer necessitate a comprehensive reassessment of the environmental, economic, historical, social, and political dimensions of excrement management. Additionally, cultural, gender, and religious influences, along with the evolutionary role of genetic factors—such as the innate disgust response—shape human perceptions, emotions, and behaviors related to excrement, toilet usage, and cleanliness, presenting further challenges to sustainable excrement management [4, 102, 116–118, 271].

Fecal contamination presents significant challenges to public health and the environment, particularly concerning surface waters. A key obstacle encountered by environmental regulatory agencies and researchers involves identifying contamination sources. Positive detections of *E. coli* or *Enterococcus* spp. may arise from various sources, including municipal and residential waste handling spills, agricultural runoff from manure-treated fields or leaks from manure lagoons, and wildlife contamination. Assessing contamination levels can be achieved through standard culturing protocols, qPCR, and the identification of human gut associated microorganisms such as *Bifidobacterium adolescentis*, *Bifidobacterium dentium*, and human enterovirus, which strongly indicate human fecal contamination [272, 273].

qPCR quantifies specific DNA sequences within a sample, supporting applications such as pathogen detection, genetic testing (ie. for mutations, allelic variations, etc.), microbial quantification, or forensic analysis. US EPA regulations outline qPCR protocols to identify fecal indicator bacteria in recreational waters such as *Enterococcus, Bacteroidales, E. coli* and *Clostridium* spp. [274–277]. Compared to culture based methods, qPCR offers a rapid turnaround time, delivering results within a few hours as opposed to 24-48 hours, thereby allowing for timely decision-making, however limited information exists when comparing *E. coli* occurrence determined by culturing and EPA Draft Method C using qPCR [275, 278]. A theoretical evaluation suggested that Method C results may align with culture-based outcomes over 90% of the time but further research is needed to assess its full compatibility [275, 278].

When applied to soils or compost, qPCR may be inhibited by molecules binding to or degrading DNA

polymerase, hindering primer extension, or interfering with fluorescence [279, 280]. However, Ibekwe and Grieve (2003) [281] demonstrated reliable quantification of *E. coli* O157:H7 cell counts in various environmental samples. Additionally, Oliver et al. (2016) [282] investigated the relationship between culture-based colony forming units (CFUs) and qPCR cell counts of *E. coli* in dairy manure exposed to different environmental conditions and temperature extremes revealing the potential underestimation of *E. coli* populations in livestock feces post defecation. Such discrepancies could pose significant challenges for monitoring water and soil contamination, particularly in the case of runoff.

Piceno et al. (2017) [54] utilized a PhyloChip microarray to assess the effectiveness of thermophilic composting in reducing fecal microbiota and pathogenic organisms below detectable limits at a HEC facility in Haiti operated by the non-profit organization SOIL. This organization collects HE and produces bagged compost for donation or sale to the community, underscoring the importance of reducing exposure to fecal bacteria and pathogens, particularly in high-risk areas. The results demonstrated that typical fecal bacteria such as *Prevotella, Lachnospiraceae*, and *Escherichia*, as well as opportunistic pathogens like *Salmonella* spp., were reduced below detection limits. Despite an overall decrease in microbial richness and abundance during the thermophilic phase, the genera *Thermobifida, Bacillus,* and *Geobacillus* increased in abundance, consistent with findings from other composting studies [214, 215, 225, 227, 228, 250].

## Night soil

In many regions globally, the use of night soil compost (NSC) remains in practice in agricultural settings [49, 50], however, concerns persist regarding its application to soils. Prior to their application, thermophilic temperatures during composting can lead to a significant reduction in opportunistic pathogens including *Prevotella, Erysipelotrichaceae, Escherichia,* and *Shigella* [54]. However, in areas where dry toilets are used, these temperatures are usually not achieved [66].

Despite challenges with managing the HE, dry composting toilets offer extremely compelling benefits. They provide accessibility for those without toilets, they effectively reduce odor through dehydration, and they are straight-forward to install due to their minimal infrastructure requirements. They are also easy to maintain, and can create economic opportunities (for those installing these systems, and for those collecting the material for reuse). Additionally, when the HE is properly managed (e.g., by composting) they can help prevent contamination of water and soils [52].

Borker et al. (2022) [49] assessed the microbial community dynamics, potential pathogenic risks, and phytotoxicity of NSC from the northwestern region of Himalayas using the V3-V4 region of the 16S rRNA gene for analysis. Their findings indicate that the NSC produced met the U.S. EPA criteria for class A compost, exhibiting safe levels of pathogens. Major genera found in the NSC included *Teredinibacter*, associated with cellulose degradation and nitrogen fixation [283], *Mucilaginibacter* posessing plant-growth-promoting traits in the rhizosphere of cotton [284], and *Haloferula*, a halophile [285]. Although ARGs were detected in bacterial strains, raising concerns for NSC reuse, the finished product was deemed non-harmful towards plant roots. This study also revealed that NSC shared predominant phyla when compared to HE, cattle manure, food waste compost, vermicompost, and activated sludge, but at the genus level NSC was notably distinct from these groups.

Similarly, Williams et al. (2018) [286] applied 16S sequencing to samples collected over a 30-day period from a microflush CT designed for the Ghana Sustainable Aid Project (GSAP). Initially, the bacterial community resembled that of the human gut, however, by day 4, the microbiome of the filter-digester bed began transitioning towards taxa commonly associated with environmental sources. The liquid effluent from the toilet was treated with solar disinfection (SODIS) and slow sand filtration (SSF). SSF appeared to introduce new taxa, possibly originating from the biological film component of the soils. Additionally, Andreev et al. (2017) [287] applied lactic acid fermentation in conjunction with thermophilic temperatures to reduce pathogen levels in HE treatment. Soils amended with the resultant compost increased the germination index for radishes and produced larger fruits in tomato plants.

Unlike Borker et al. (2022) [49], a study of dry composting toilets in Mexico evaluated the effectiveness in reducing fecal coliforms, with only 36% of toilets achieving class A compost status after 6 months. In this study, desiccation was the primary mechanism for fecal coliform reduction rather than biodegradation. Drier samples with greater solar exposure resulted in a higher proportion of class A samples [149]. Conversely, Gao et al. 2021 [288] found that the proportion of potential pathogenic bacteria in dry toilets was higher than that in septic tanks used in rural China, suggesting that septic tanks were better than dry toilets in reducing HE pathogen load.

In a study conducted in the Kathmandu Valley, Nepal, it was found that *Clostridium perfringens* exhibited the highest resistance among bacterial indicators in urine-diverting dehydrating toilet units, where wood ash was a primary additive [289]. Due to elevated pH from the addition of ash, microbial degradation rates and therefore heat production decreased [143]. When environmental conditions are unfavorable, *C. perfringens* forms endospores [290] and its presence may be attributed to the existence of small microhabitats harboring favorable environmental conditions, such as pockets of high nutrient or oxygen availability. The failure of these toilets to attain thermophilic temperatures may have also fostered regrowth.

In a different study, Borker et al. (2020) [291] identified the psychrotrophic bacterium *Glutamicibacter arilaitensis* LJH19. Through whole-genome sequencing, they identified its enzymatic capabilities, demonstrating amylase, cellulase, and xylanase activities even at a low temperature of 10°C. Furthermore, this organism exhibited plant growth promoting (PGP) characteristics and evidence of siderophore production, a trait with potential applications in various fields including bioremediation, biocontrol of plant pathogens, and the development of novel antimicrobial therapies [292].

## Pit latrines

A significant proportion of the global population depends on pit latrines for sanitation [60], and sampling these pits for waste-based epidemiology presents a substantial occupational hazard to public health professionals. In a study by Capone et al. (2021) [293], various depths of pit latrines were investigated to identify the most effective method for assessing pathogen levels relevant to waste-based epidemiology. The findings revealed no significant differences in pathogen detection across the different depths examined. Consequently, the authors proposed that surface sampling of latrines could yield representative data, offering valuable insights into community health while minimizing exposure risks for public health professionals. Moreover, these findings could inform the optimization of mass vaccination campaigns and the enhancement of water, sanitation, and hygiene (WASH) interventions, which are essential for preventing waterborne diseases and improving public health outcomes [294].

Smith et al. (2023) [191] analyzed the same pit latrine dataset of Capone et al. (2021) [293] to characterize the microbial populations across various depths and geographical regions in Malawi. Their analysis revealed a dominance of fermentative bacteria, notably *Clostridium sensu stricto 1*, across all depths and locations, with taxa associated with methanogenesis pathways exhibiting increased prevalence with greater depth. The microbiomes in pit latrines were distinct from those found in activated sludge, municipal sewage sludge anaerobic digesters, and the human gut. These findings suggest that the microbial community structure and metabolic processes in pit latrines are robust and likely driven by similar environmental conditions rather than geographic location or waste characteristics. The microbial processes in pit latrines may also be specialized for the specific conditions found there, such as varying oxygen levels, nutrient availability, and waste composition. Understanding these distinctions is crucial for optimizing waste management practices, as strategies effective in activated sludge or anaerobic digesters may not be directly applicable to pit latrines, and vice versa. This could influence the design of interventions aimed at improving sanitation and reducing associated health risks. We hypothesize that microbiome change over time, and microbial succession transitioning from feces-like microbiomes to environmental microbiomes, occurs at a far slower pace in pit latrines than in effectively managed composting toilet systems such as those in other studies [49, 54, 286]. Including a temporal component in future studies of pit latrine microbiomes would enable testing of this hypothesis, and support comparisons of safety of the two approaches from a microbial ecosystem perspective.

While some investigations have raised concerns about potential health hazards to communities, others have demonstrated that, under specific conditions, the final product of pit latrines can be pathogen-free and beneficial for agricultural purposes, enhancing soil quality and promoting higher agricultural yields. Variations in individual exposure levels stem from factors such as pit emptying practices, sludge storage methods, and the adoption of protective measures [61]. In regions where the parasitic worm *Ascaris lumbricoides* is prevalent, sodium hypochlorite-based disinfectants proved most successful, compared to carbolic acid-based ones, to inactivate *Ascaris* eggs during pit emptying [62]. Individuals storing sludge in sacks and using personal protective equipment (PPE) during handling and application of the sludge experienced lower exposure through inhalation, ingestion, and dermal contact [61].

### Co-composting

Co-composting HE with food and landscaping waste presents a viable solution that also reduces the volume of compostables deposited in landfills [64]. However, the risks of pathogen exposure are still concerning if not properly treated. Resource-oriented sanitation (ROS) systems treat HE as a valuable resource capable of generating energy, fertilizers, and soil amendments. Hashemi et al. (2019) [295] investigated the effectiveness of incorporating sawdust, rice husk, and rice husk charcoal, in the composting process of fresh feces within ROS systems. The addition of rice husk charcoal significantly enhanced compost properties and resulted in the elimination of *E. coli* strains after five weeks of composting. This suggests that rice husk charcoal can serve as an effective bulking material in ROS systems, providing a low-cost, nature-based treatment option for individuals who have access to this material [295].

# Application of Multi-omics Tools to Further Improve Sustainable Waste Management

Results from qPCR and amplicon sequencing have revealed discernible patterns of microbial succession, elucidated effective pathogen reduction strategies, and identified specific microbial taxa associated with inoculation, competition, PGP, and ARG reduction. However, more targeted approaches such as metagenomics, metatranscriptomics, and metaproteomics provide deeper insights into the intricate dynamics of these ecosystems.

These advanced techniques are increasingly cost-effective and accessible, providing valuable insights into enzymatic biodegradation, pathogen detection and identification, ARG detection and identification, and metabolic activities and pathways, providing a promising direction for advancing sustainable waste management. This section reviews the limited studies applying other -omics technologies to HEC and highlights their use in other waste management systems to showcase the potential benefits for HEC research.

During composting, cellulose, the most abundant carbohydrate in plant biomass, serves as a vital carbon source. A metaproteomics analysis of agricultural waste mixed with cow manure demonstrated that bacterial communities, dominated by *Thermobifida* appeared to be primarily responsible for cellulose degradation, and that fungal communities, dominated by *Thermomyces* and *Aspergillus*, were associated with hemicellulose degradation [296].

β-glucosidase (BGL) is one enzyme responsible for degrading cellulose to glucose. Liu et al. (2015) [297] found that fungi were responsible for producing BGL in municipal compost systems and that thermotolerant fungal communities did not persist into the thermophilic phase as expected. Zhang et al. (2020), [298] using metatranscriptomics, discovered that microbial communities producing BGL exhibited differential expression of glucose-tolerant and non-glucose-tolerant BGL genes throughout composting. This variation likely reflects the communities' use of multiple regulatory strategies to adapt to changing carbon-metabolizing pressures, such as fluctuating glucose concentrations during the composting process.

Liu et al. (2015) [297] identified carbohydrate metabolism as the principal metabolic pathway in municipal composting by Actinobacteria, Bacilli, Alphaproteobacteria, Betaproteobacteria and Gammaproteobacteria classes. They also found an enzyme involved in nitrogen fixation may have only been produced by *Rhodobacter sphaeroides* in their composting system, and another enzyme for denitrification may have only been produced by *Pseudomonas stutzeri*, illustrating an ability to associate microbial activity with specific microbial taxa. This knowledge could be extremely useful in the development of inoculants as the functional space of microbes in waste management systems is mapped. Additionally, the discovery of genes involved in methanogenesis [243] could be leveraged through bioengineering to improve methane gas capture from agro-industrial wastes, demonstrating potential to support renewable energy development [299].

Metagenomics has significantly enhanced our understanding of the microbial and viral communities present in various composting systems, shedding light on potential health risks associated with their use. Blomström et al. (2016) [91] used metagenomics to identify DNA viruses, primarily finding DNA bacteriophages in vermicompost, as well as the presence of pathogens and a high abundance of *Aeromonas* which can cause gastrointestinal infections in humans [300]. More recent studies using metagenomics by Liu et al. (2022) [46] and Borker et al. (2022) [49] identified the presence of antibiotic- and heavy metal-related resistance genes in animal manure and night soil compost, respectively. These investigations reinforce the environmental and human health risks associated with utilizing compost derived from sewage sludge and HE, as regulated by the USDA National Organics Program [301].

Several researchers have applied metagenomics and metatranscriptomics to better understand the microbial complexities of large-scale composting within the São Paulo Zoo Park in Brazil. Organic material included animal excrement and feed, WWT sludge, and landscaping debris. Metagenomic sequencing revealed that the decomposition of recalcitrant lignocellulose was primarily facilitated by bacterial enzymes from the Clostridiales and Actinomycetales orders [93]. The results also suggested that cellulose and hemicellulose degradation were likely performed by cellulosomal enzymes, known for their efficiency in breaking down complex plant polysaccharides. Antunes et al. (2016) [92] also applied metagenomics to the same dataset, resulting in the near-complete reconstruction of the genomes of five bacterial species associated with biomass-degrading environments. Additionally, they identified a novel biodegrading bacterial species, potentially representing a new genus within the Bacillales order.

Metatranscriptomics data from Antunes et al. (2016) [92] allowed the researchers to reclassify the composting process into three distinct phases (beginning, middle, and end of composting) based on metabolic functions. The beginning phase was associated with breaking down and utilizing easily degradable organic nutrients, the middle was associated with the degradation of more complex carbon sources, and the last phase was associated with amino acid metabolism, energy production/conversion, and signaling processes. These functional activities may provide opportunities for biomarker development to assess compost safety, and inform inoculant development to further enhance compost efficiency. More recently, Braga et al. (2021) [42] analyzed the same São Paulo Zoo Park dataset using metatranscriptomics and identified numerous active ARGs linked to *Pseudomonas thermotolerans* and *Mycobacterium hassiacum*.

The collective efforts of Martins et al. (2013), Antunes et al. (2016), and Braga et al. (2021) [42, 92, 93] have provided a comprehensive examination of microbial dynamics within the co-composting of food, landscaping, and manure at São Paulo Zoo Park. Their investigations elucidated the temporal fluctuations in the abundance of specific organisms and their functional activity, culminating in the description of twenty-four novel bacterial species. Such comprehensive investigations into composting systems serve as valuable models that could improve HEC using CTs, and demonstrate the utility and power of applying integrated multi-omics technologies to advance sustainable waste management.

# Conclusion

Our society generates vast quantities of excrement and other organic materials that are currently managed through largely linear systems but which have the potential to be cycled for improved public health, agricultural productivity and safety, and environmental sustainability. Composting is the microbial biotechnology aimed at driving that transition, and applications of modern microbiome -omics and related technologies have vast capacity to support continued advances in composting science and praxis. In this article, we have reviewed literature focused on applications of microbiome technologies to study composting systems and reactions. We conclude that work to-date has demonstrated exciting basic and applied science potential from studying compost microbiomes.

Composting toilets and human excrement composting in particular can support improved management of what is likely the most problematic organic waste material: human excrement. We have found that there are relatively few studies of these important systems that employ modern microbiome science technologies, but the studies that have been performed demonstrate the potential of this work [49, 54, 286]. We consider this an opportunity for research that can lead to technological and economic development and have impacts in basic and applied microbiome science. Expanded use of composting toilets and human excrement composting across the globe - which would be an important outcome of this research - also relies on overcoming social, cultural, political, and economic barriers, and would be a challenge that could last multiple human life spans. Like planting a tree, the best time to begin is twenty years ago, and the second best time to begin is now.

Acknowledgement of the importance of cycling nutrients from human excrement, and avoiding the use of dwindling landfill space for disposal of this material, is clear in the expanded use of biosolids in agriculture, forest restoration, and other applications over the last five decades [218, 302]. However there is also growing recognition of the limitations of biosolids: namely that highly toxic materials, such as PFAS, are introduced as a result of human excrement having been combined with industrial wastes during collection [17]. Composting of human excrement before it is combined with other waste streams, rather than after, holds a key to safer cycling of these materials, and microbiome technologies can be pivotal to further developing the use of human excrement composting (Meilander et al. (2024) - *Microbiome Journal*, in press).

qPCR, amplicon sequencing, and other -omics technologies have revolutionized our ability to discover and identify specific organisms, sequence their genomes, and identify key genes and proteins influencing microbial ecology. By better understanding the development and activities of microbial communities under certain conditions, including their capacity to degrade toxins such as PFAS, bacterial, fungal, and viral pathogens, and pharmaceuticals, we can further enhance the performance and safety of composting systems. This opens avenues for targeted inoculation development, biomarker discovery, and other bioengineering applications with economic potential.

Linear waste management systems are unsustainable, and a shift towards nutrient cycling, reduced water and energy consumption, and mitigation of pathogen and toxin exposure is urgently needed. By leveraging microbiome science technologies, numerous stakeholders — including compost managers, researchers, policymakers, for-profit businesses, non-profit entities, and humanitarian aid groups striving to achieve the United Nations' Sustainable Development Goal 6 — can spearhead the development and adoption of more sustainable waste management practices.

# Figures and tables

**Figure 1. Composting phases.** The flow chart illustrates the four distinct phases of composting, each characterized by specific temperature ranges and processes. Composting is commonly divided into phases to better understand the biological and chemical transformations that occur during decomposition, where phases are defined based on temperature changes. [34]. Created in BioRender.

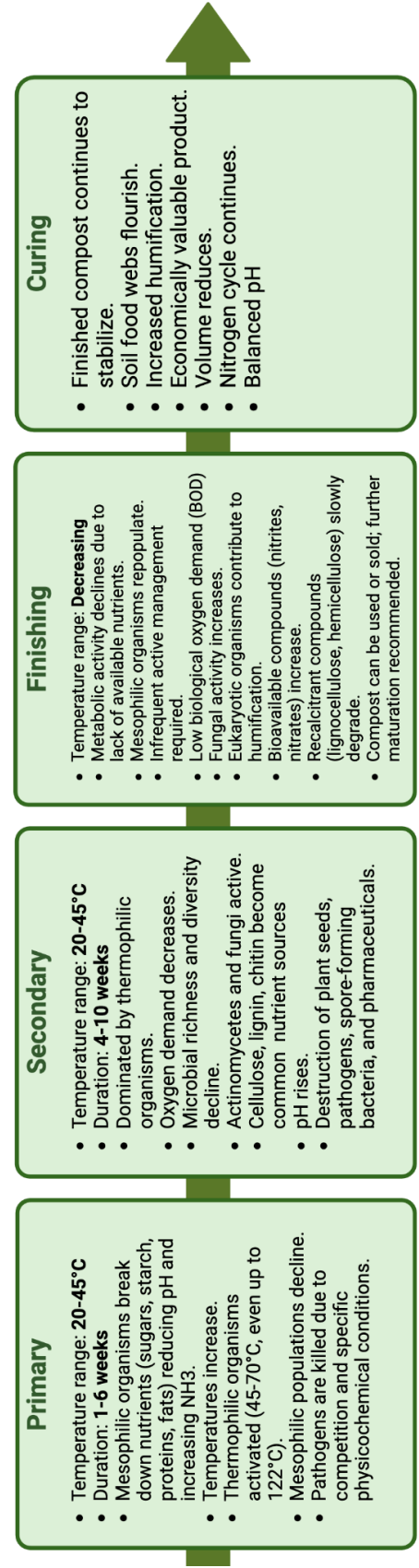

**Table 1. Optimal Parameters and Key Variables for Effective Composting.** A summary of the essential compost variables and their preferred ranges for achieving effective composting. Each variable plays a critical role in the composting process. Adapted from [27].

| Variable | Description | Preferred Range |
|---|---|---|
| Carbon to Nitrogen Ratio (C:N) | Ratio of carbon to nitrogen (C:N) in compost feedstock. | 25:1-40:1 |
| Moisture | Water content in the compost pile. | 50%-60% |
| Oxygen Content | Adequate oxygen availability within pore spaces for aerobic decomposition. Proper aeration helps maintain oxygen. | > 10% |
| Temperature | Optimal temperatures to achieve proper sanitization. | 50-65°C |
| pH | Acidity or alkalinity of the compost. | 6.5-8.0 |
| Bulk Density | The weight of compost per unit volume - maintains proper aeration and porosity. | 400-600 kg/m$^3$ |
| Feedstocks | Organic materials collected for composting (e.g., kitchen scraps, landscaping waste, manure). | |
| Bulking Materials | Added to feedstocks to optimize microbial activity and improve compost texture. | Added with feedstocks to achieve preferred ranges |
| Qualitative Variables | Color, odor, and texture serve as quick quality indicators. Attention to other variables is crucial. | |